\documentclass{elsart}
\usepackage{graphics}
\usepackage{amsmath}
\begin{document}
\begin{frontmatter}
\title{Adsorbate Interactions of CO Chemisorbed on Pt(111)}
\author{R. Brako\thanksref{Corr}} and
\author{D. \v{S}ok\v{c}evi\'{c}}
\address{
Rudjer Bo\v{s}kovi\'{c} Institute, P.O.~Box 1016, 10001 Zagreb, 
Croatia }
\thanks[Corr]{Corresponding author. e-mail:
radovan.brako@irb.hr}

\begin{abstract} 
We show that the observed repulsive interaction between
CO molecules on the Pt(111) surface can be explained by
the coupling of the Pt--CO separation with Pt--Pt coordinates
in the substrate.
The observed long range of the interaction and the 
non-monotonic distance dependence are reproduced.
The magnitude of the multiphonon decay of the Pt--CO
vibration calculated in this model is also in agreement
with experiment. \\


\end{abstract}

\end{frontmatter}

Adsorbate-adsorbate interactions are an essential aspect
of potential energy surfaces in chemisorption on metal surfaces.
They manifest in effects such as the formation of adsorbate 
islands and regular structures, they influence the shape
of thermal desorption spectra, they may be very 
important for diffusion and reactions between adsorbates, etc.

The properties of the clean Pt(111) surface and of
CO chemisorption are complex. There is
large tensile stress within the first atomic layer of the 
clean Pt(111) surface (Ref. \cite{Hohage95} and references therein,
\cite{Boisvert97})
This is caused by an abrupt change in the electronic structure at
the surface, with a consequent decrease of 
the optimum distance between the atoms in the 
first layer compared with the bulk lattice constant.
The effect is not strong enough to induce
reconstruction at room temperature, 
unlike the Pt(100) surface which reconstructs
by forming a dense hexagonal layer
\cite{Heilmann79,Behm86} (although a 
non-reconstructed metastable Pt(100) surface can also be 
obtained under certain conditions). 
The reconstruction of the Pt(111) surface does occur at high 
temperatures in the presence of saturated Pt vapour \cite{Hohage95}, 
where a network of linear structures consisting of more
densely packed atoms
develops. Interestingly, the distance of the first to the
second atomic layer is the same as in the bulk or even
slightly larger, as both experimental and theoretical
evidence suggests (Ref. \cite{Boisvert97} and references therein).
 However, at step
edges the interatomic distances contract \cite{Boisvert97,Hammer97}.
CO molecules adsorb initially on the on-top sites, but the
energy difference for the adsorption into bridge sites is
obviously rather small, so that some bridge adsorbates are
found already at coverages above 0.15 \cite{Yoshinobu96}. Several
regular adsorbate structures have been observed \cite{Tushaus87}.
At a coverage of 0.29 a complex regular structure of on-top adsorbates
is formed, consisting of hexagonal patches of coverage 1/3
separated by unoccupied atoms.
According to Ref. \cite{Tushaus87}, this is the densest structure of
on-top adsorbates only, and further chemisorption occurs into
bridge sites, but other authors claim that a regular 
$(\sqrt{3} \times \sqrt{3}){\rm R}30^{\circ}$ structure at a coverage of 1/3
exists (Ref. \cite{Yeo97-King} and references therein).
The regular structure at 0.5 contains
an equal number of on-top and bridge adsorbates. 

The existence of regular adsorbate structures is indicative of
strong interactions between the adsorbates. Furthermore,
the adsorption energy decreases from around 1.9 eV at low 
coverages to 1.66 eV at coverage 0.33 and
 to 1.2 eV at half coverage \cite{Yeo97-King}, which means
that the interaction is repulsive. There is evidence 
that repulsive forces exist between adsorbates several 
lattice constants apart, and that they depend non-monotonically
upon distance \cite{Skelton94}. Another indication of the importance 
of adsorbate-adsorbate interactions is the peculiar coverage 
dependence of the linewidth of the CO--Pt vibrations of the on-top
species, which has been observed by IR spectroscopy \cite{Persson89}. 

Theoretical work on the adsorbate-adsorbate interaction has been
inconclusive, apart from the general agreement that the
interaction is mediated through the substrate. 
First-principle electronic calculations
have failed to clearly identify the interactions,
or have given attractive interactions, which is not 
in agreement with experiment \cite{Hammer97,Jennison96}.
However, such calculations have not included full relaxation of
the substrate atoms upon chemisorption, either because the 
lateral relaxation of surface atoms is precluded by
symmetry when periodic geometries are used, or because
there are too many degrees of freedom to investigate 
relaxations and there is large intrinsic stress between
substrate atoms in the case of cluster geometries.

In this paper we show that interaction energies for on-top
chemisorption of CO on Pt(111), which are in agreement
with experimental data, can be obtained from a model based on
lateral relaxation of Pt atoms around the adsorption site,
with the same distance dependence as the interaction potential.

We describe the adsorbate--substrate interaction by the 
Morse potential
\begin{equation}
V(z - z_0) = E_0 (e^{- 2 \alpha (z - z_0)} - 2e^{-\alpha (z - z_0)} ),
\label{Morse}
\end{equation}
where $E_0$ is the adsorption energy, $z$ the coordinate of the CO
molecule and $z_0$ the coordinate of the Pt atom. (In our
notation, the minimum of the potential is at $z-z_0=0$. 
It would be more consistent to explicitly include the 
equilibrium distance $z_0$. For simplicity,
here and in the following we omit the equilibrium distances between
atoms.) The constant $\alpha$ can be expressed as
\begin{equation}
\alpha = \sqrt{\frac{\mu \Omega ^2}{2 E_0}},
\label{alpha}
\end{equation}
where $\mu$ is the reduced mass (assuming, say, a free Pt--CO complex), 
and $\Omega$ the frequency of Pt--CO vibrations, which are 
known quantities. If we expand the potential (\ref{Morse}) around the
equilibrium, the terms linear in $(z - z_0)$ cancel, the quadratic
terms give the harmonic force and higher terms give non-harmonicity.
This means that by using the Morse potential we obtain an estimate
(usually quite reliable) of non-harmonicity in terms of
known quantities.

We extend this approach by including into the potential
terms which couple the Pt--CO coordinate $(z - z_0)$ with other
interatomic distances in the substrate, with the same 
exponential dependence as the {\it attractive} part of the 
Morse potential:
\begin{equation}
c e^{-\alpha (z - z_0)}(x_0 - x_1). \label{nondiagonal}
\end{equation}
We include only the attractive part of the 
Morse potential because it corresponds (more or less) to 
the formation of the chemical bond by the valence electrons of
the adsorbate and the substrate, which is expected 
to modify the bonding of the substrate atom to its neighbours
as well.
The repulsive part of the potential, on the other hand, is due
mostly to the overlap of the inner (chemically inert) orbitals of
the adsorbate and the substrate atom, which does not
affect other bonds.

Eq. (\ref{nondiagonal}) expands around the equilibrium point into
\begin{equation}
c (x_0 - x_1) - c \alpha (z - z_0)(x_0 - x_1) + \ldots .
\label{nondiagonalexpansion}
\end{equation}
When added to the harmonic potential which describes the
forces between other substrate atoms, the first (linear) term gives 
a downward shift of the minimum of the potential,
i.e. a relaxation energy, and a change
of the equilibrium distance in the $(x - x_0)$ coordinate, while the
second (non-diagonal quadratic) term couples the 
motion of the $(x - x_0)$ and the $(z - z_0)$ coordinates. 
It has recently been argued that 
the non-diagonal term is large on the ground that it allows the
multiphonon decay of Pt--CO vibrations \cite{Brako95},
which would otherwise
have very small probability for on-top adsorbates. 
We shall return to this problem further on.

\begin{figure}[htb]
\begin{center}
\resizebox{9cm}{!}{\includegraphics{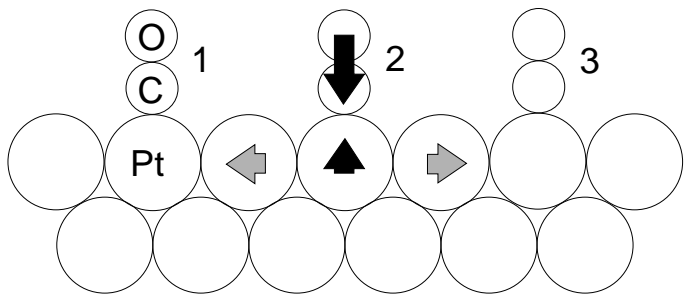}}
\end{center}
\caption{A schematic picture of CO adsorption. At a low coverage
(CO molecule at position 2 only) the adjacent Pt atoms shift
to a relaxed position, and participate in Pt--CO
vibrations (grey arrows). At high coverages (positions 1 and 3
also occupied), the Pt atoms in between the adsorption sites can
neither relax laterally, nor can they participate in 
vibrations if all CO adsorbates vibrate in phase.}
\label{fig:scheme}
\end{figure}

The effects of the linear term may be significant or not,
depending upon the nature of the $(x - x_0)$ coordinate. If 
$(x - x_0)$ is a distance to a Pt atom in the second layer, 
its relaxation, although important in principle, is difficult to
determine experimentally, while the energy gain is small compared
with the total depth of the adsorption well. The situation is different
if $(x - x_0)$ is the distance to another Pt atom in the first 
layer, i.e. a coordinate parallel to the surface. 
If the CO coverage is small, each adorbed molecule 
may be treated independently, and more or less the
same reasoning applies. However, if the adsorbate concentration is
large, the presence of another CO on the second-nearest neighbour
will inhibit the relaxation of the middle Pt atom in the
direction parallel to the surface, as shown in Fig. \ref{fig:scheme}.
This means that at large CO coverages there will be no energy 
gain owing to relaxation, and the total adsorption energy in a dense 
overlayer will be {\it smaller} than for an isolated CO, i.e. 
there is a net repulsive interaction between adsorbates. In
the following we develop a quantitative theory of the
effect.

We conclude these introductory remarks by noting that the
{\it quantitative}
aspect of the described approach has all virtues and problems 
inherent to the use of the Morse potential for describing 
chemisorption.  Ultimately,
only first-principle calculations can 
show how large the deviations of a more realistic potential from
the Morse function are.

In our model we include only the first atomic layer of the Pt(111)
surface, with one or two CO molecules adsorbed on it. We consider
the motion perpendicular to the surface ($z$-direction) only
for the Pt--CO complex (coordinates $z_0$ and $z$; the C--O bond is
assumed rigid), and only the motion parallel to the
surface of the other Pt atoms (coordinates  $\vec{r}_i=(x_i,y_i)$).
The complete
Hamiltonian in the harmonic approximation for one adsorbate is
\begin{equation}
H = T + V, 
\label{Hamiltonian}
\end{equation}
where $T$ is the kinetic energy:
\begin{equation}
T = {1 \over 2} m \dot{z}^2 + {1 \over 2} M \dot{z}_0^2 +
 {1 \over 2} \sum_{i} M \dot{\vec{r}}_i^2
\label{kinetic}
\end{equation}
and $V$ is the potential energy:
\begin{eqnarray}
V & = & {1 \over 2}Kz_0^2 + {1 \over 2}k(z - z_0)^2 + {1 \over 2}
 \sum_{i} \sum_{j=1}^{6} {1 \over 2}K_1 [\hat{{\vec{r}}}_{ij} \cdot 
 ({\vec{r}}_i - {\vec{r}}_j )]^2 + {1 \over 2} \sum_{i}K_2 {\vec{r}}_i^2
  \nonumber \\ 
 & & + \sum_{j=1}^6 k_1 (z - z_0)\hat{\vec{r}}_{0j}\cdot({\vec{r}}_0
  - {\vec{r}}_j)
 - \sum_{j=1}^6{k_1 \over \alpha} \hat{\vec{r}}_{0j}\cdot ({\vec{r}}_0 -
 {\vec{r}}_j).
 \hfill
\label{potential}
\end{eqnarray}
Here the index $i$ runs over surface atoms, $i=0$ being the adsorption
site, the index $j$ runs over the six Pt atoms around the atom $i$, and
$\hat{\vec{r}}_{ij}$ is a unit vector along the direction from $i$ to $j$.
The Hamiltonian for two CO adsorbates is obtained by introducing
another set of $z$-dependent terms at another site $i$.

The kinetic energy terms are self-explanatory. The potential energy
terms with the force constants $K_1$ and $k$ describe, respectively, central
harmonic forces between neighbouring atoms in the $x$--$y$ plane, and
between a Pt atom and the CO molecule
in the $z$ direction. The term with a force constant $K$
is the coupling of the Pt--CO complex to a rigid `hard wall' which
describes the Pt bulk, and the term with $K_2$ couples the $x$--$y$ motion
of Pt atoms to their equilibrium positions, i.e. to the atoms in the
second layer, which are again assumed to be rigid. The last two terms
with a force constant $k_1$
are the off-diagonal coupling of the Pt--CO bond to the surrounding
Pt atoms, as explained in deriving Eqs. (\ref{nondiagonal})
and (\ref{nondiagonalexpansion}).

It is necessary to include the $K_2$ term, because the model with
forces only between the atoms in the first layer
(i.e. the $K_1$ term) gives a
response which is too soft, since it only has an acoustic branch 
of vibrations in the $x$--$y$ plane, whereas
most of the density of states for motion
in the $x$--$y$ plane on a real surface is at finite frequencies even at
small wavevectors. The masses $m$ and $M$ in Eq.~(\ref{kinetic}) are
those of CO and Pt, respectively.
The values of the force constants $K_1=0.65\times  10^5$ dyn/cm,
$K_2=0.2\times 10^5$ dyn/cm, and $k=3.05\times 10^5$ dyn/cm
are chosen so as to reproduce the frequency of Pt--CO
vibrations ($\sim 58$ meV \cite{Persson89})
and typical phonon frequencies (from $\sim 6$ meV to
$\sim 24$ meV \cite{Bortolani89,Schweizer89}),
and $\alpha$ is
given by Eq. (\ref{alpha}). This is a sufficient minimal model for
our problem, because we need the response of surface
atoms to an external force, i.e. an average quantity, 
rather than the complete
vibrational spectrum, although it would be a poor description
of surface phonons for most other purposes.

\begin{figure}[htb]
\begin{center}
\resizebox{9cm}{!}{\includegraphics{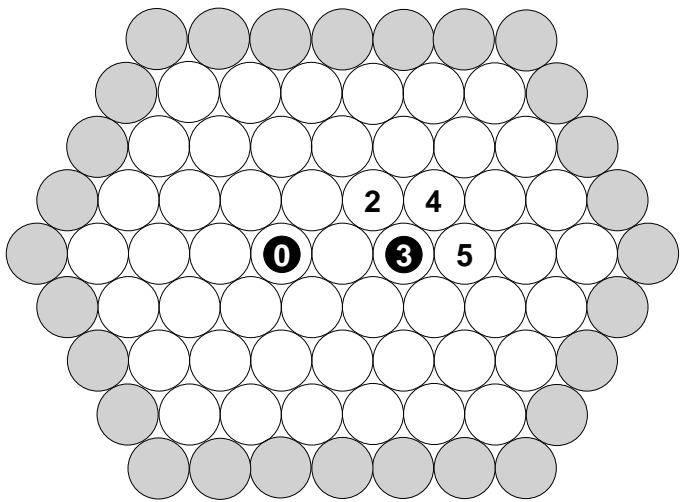}}
\end{center}
\caption{The geometry of Pt clusters used in the calculations,
for CO adsorption on third nearest-neighbour sites (black
circles). The outermost Pt atoms (grey) are kept fixed,
whereas others are allowed to relax. Other adsorption sites for
which calculations have been performed are indicated by numbers.
The shapes of the clusters are adjusted accordingly.}
\label{fig:cluster}
\end{figure}

We are now in a position to calculate the relaxation energies
by finding the minimum of the potential energy $V$, 
Eq. (\ref{potential}). We performed the calculations for a single
adsorbed CO molecule, and for two CO molecules adsorbed on, in turn,
second, third, fourth and fifth
nearest-neighbour sites. We did not consider first nearest-neighbour
adsorbates, since these are not observed in experiment. We used
finite clusters of Pt atoms, as shown in
Fig. \ref{fig:cluster} for the
third nearest-neighbour case, keeping the outermost Pt atoms (grey)
fixed. Full convergence of the relaxation energy was
obtained using about ten layers of Pt atoms around the adsorption
site. As already mentioned, the relaxation is incomplete when the
adsorbates are nearby, since the two of them act on
one or more Pt atoms with forces in opposite directions.
We denote the relaxation energy for a single adsorbate by
$E_0$, for two atoms on second nearest-neighbour sites by
$E_{\rm 2NN}$, etc.
The interaction energy is defined as $W_{\rm 2NN}=2E_0-E_{\rm 2NN}$, i.e.
the difference in energy when the two adsorbates are on
second nearest-neighbour sites and when they are far apart.

In Table \ref{tab:energies} we give the calculated
values of the interaction energies.
We have chosen the value of the only remaining parameter in the model,
the off-diagonal force constant which couples the Pt--CO and Pt--Pt
coordinate, $k_1=0.35\times 10^5$ dyn/cm,
so that good agreement with experimental values is obtained.
To our knowledge,
the only other value proposed for this quantity is
$k_1 = 0.141\times  10^5$ dyn/cm,
as reported in Ref. \cite{Lahee86}. It was used
in calculating the eigenfrequencies of various modes of CO/Pt(111)
vibrations, in particular hindered translations, observed 
experimentally in helium scattering experiments.
However, this force constant in
Ref. \cite{Lahee86} was of minor importance,
as shown by the fact that it was taken over unchanged from
a study of vibrations of CO on Ni(100) \cite{Richardson79},
where, in turn, it
was assumed to be equal to the value of the \mbox{CO--Pt--CO} force
constant of nickel carbonyl. Thus it obviously cannot be considered
a reliable estimate for the CO/Pt(111) system.
\begin{table}
\caption{Interaction energies, in eV. $W_{\rm 2NN}=2 E_0-E_{\rm 2NN}$,
  where $E_0$ is the relaxation energy of a single CO molecule,
 $E_{\rm 2NN}$ is the relaxation energy for two CO molecules
  adsorbed at second nearest-neighbour sites, etc.}
\label{tab:energies}
\begin{tabular*}{\textwidth}{l@{\extracolsep{\fill}}lllll} \hline
 & $W_{2\rm NN}$ & $W_{\rm 3NN}$ & $W_{\rm 4NN}$ & $W_{\rm 5NN}$
        & $E_0/6 $ \\ \hline
Theory & 0.015 & 0.031 & 0.012 & 0.014 & 0.051 \\
Experiment  & 0.010\footnotemark[1]
  & 0.034\footnotemark[1] & 0.020\footnotemark[1]
  & & 0.039\footnotemark[2] \\ \hline
\end{tabular*}%
\\
\footnotemark[1]{\footnotesize Ref. \cite{Skelton94}, low CO coverage.}\\
\footnotemark[2]{\footnotesize Ref. \cite{Yeo97-King}, CO coverage of 0.33.}
\end{table}

We have compared our results for the interaction energies with two 
published sets of data. The values in Ref. \cite{Skelton94} were
obtained by analysing the experimental adsorption isotherms
using a transfer-matrix approach. They are for low coverages
of adsorbed CO, up to 10\%. The value in Ref. \cite{Yeo97-King}
is obtained from the change of the adsorption energy between a low
coverage and a coverage of 0.33, which corresponds to a regular 
$(\sqrt{3}\times\sqrt{3}){\rm R}30^{\circ}$ structure of
on-top adsorbed CO molecules.
Since in this structure no lateral relaxation of Pt atoms is 
possible, we compare this interaction energy with one-sixth 
of our relaxation energy for a single CO.  

Overall, the quantitative agreement of our results with experiment
is good. We have obtained the same non-monotonic dependence
of the interaction
energies as in Ref. \cite{Skelton94}. The repulsive interaction is
particularly large when the two adsorption sites lie along a row
of Pt atoms, i.e. in the case of third and fifth nearest neighbours.
In Ref. \cite{Skelton94} the fifth nearest-neighbour site is not
taken into account, but an accordingly larger interaction between
the fourth nearest-neighbour adorbates is obtained.
We have calculated that the relaxation of the position of the
six surrounding Pt atoms for an isolated CO adsorbate is
$\Delta x = 0.07$~{\AA}, or about 3\% of the nearest-neigbour
Pt--Pt distance.
There are no experimental
data to compare with, but this value is similar to the calculated
relaxation of Pt atoms on step edges upon CO chemisorption
\cite{Hammer97}. Also, the change in Pt--CO distance
obtained, 0.05~{\AA},
is close to the value calculated in Ref. \cite{Hammer96}.

Next, we briefly address the multiphonon decay of Pt--CO 
vibrations. The present approach has already
been used in Ref. \cite{Brako95}, but we are now able
to obtain a more quantitative estimate.
As the energy of Pt--CO vibrations at 58 meV is high
above the largest energy of Pt phonons, at about 24 meV,
the decay can occur only via anharmonicity-mediated
three-phonon (or higher order) processes. Assuming that the 
dominant anharmonicity is that of the \mbox{Pt--CO} bond, the anharmonic
terms can be found expanding the Morse potential (\ref{Morse})
beyond the harmonic approximation. Since these terms are
small, the multiphonon transitions can be evaluated 
perturbatively after the harmonic Hamiltonian (\ref{Hamiltonian})
is diagonalized.

We denote the initial coordinates by a vector
${\sf{x}} = (z,z_0,x_i,y_i)$,
the transformed ones by ${\sf{v}} = (u,v_i)$, the mass matrix 
corresponding to the kinetic energy (\ref{kinetic}) by $\sf{T}$
and the orthogonal transformation matrix by $\sf{A}$:
\begin{equation}
{\sf{v}} = {\sf{A}}^\dagger {\sf{T}}^{1/2} \sf{x}.
\end{equation}
The matrix $\sf{A}$ transforms the Hamiltonian
(\ref{Hamiltonian}) into the diagonal form
\begin{equation}
H = {1 \over 2} \dot{u}^2 + {1 \over 2}\Omega^2 u^2 +
 \sum_{i}\left({1 \over 2}
\dot{v}_i^2 + {1 \over 2}\omega_i^2 v_i^2 \right),
\end{equation}
where $\Omega$ is the large frequency corresponding to the CO--Pt
vibration and $\omega_i$ are smaller `phonon' frequencies.
We express the Pt--CO coordinate in terms of the diagonalized
modes:
\begin{equation}
z - z_0 = g u - \sum_{i} h_i v_i.
\label{molecular}
\end{equation}
The coefficient $g$ is large, whereas $h_i$ are small, 
expressing the fact that the phonon modes which couple
to Pt--CO vibrations have only a small admixture of the
Pt--CO coordinate.

We now insert (\ref{molecular}) into the anharmonic terms
$(z - z_0)^3, (z - z_0)^4$, etc.,
obtained from the expansion of the Morse potential, and obtain
products of various powers of $gu$ and $h_iv_i$. The dominant 
contribution to the vibrational decay comes from three-phonon
processes, i.e. from transition probabilities from an initial
state $|i\rangle $ consisting of a singly excited $u$ (Pt--CO)
vibration to a final state $|f\rangle$ with $u$ in the ground  
state and three excited phonon states $v_i$ such that the total 
energy is conserved.  
The transition probability is
\begin{equation}
 P_{if} = {{2\pi}\over{\hbar}}\left|\langle i|H'|f \rangle\right|^2
 \delta\left(E_i - E_f \right),
\end{equation}
where $H'$ is the anharmonic part of the Morse potential (\ref{Morse})
(essentially $uv_1v_2v_3$), and the total decay width is
\begin{equation}
\Gamma = \hbar \sum_{f} P_{if}.
\end{equation}
According to (\ref{molecular}), $P_{if}$ is proportional to
$(gh_1h_2h_3)^2$. In the case
of no off-diagonal coupling, $k_1=0$, in our model there is only one
$h$ and, to leading order, the coefficients are
\begin{equation}
 g = {1 \over{\sqrt{m}} }, \nonumber
\end{equation}
\begin{equation}
h = {1 \over\sqrt{M}} \left({\omega \over \Omega}\right)^2.
\label{reduction}
\end{equation}
(Of course, the energy conservation cannot in general be satisfied 
in the model with only one phonon coordinate $v$, but the same
conclusion holds if the model is generalized introducing a 
projected phonon density of states $\rho(\omega)$, as in Ref.
\cite{Brako95}.) The frequency ratio in (\ref{reduction})
is the `reduction factor', 
which makes the decay probability too small by about five orders
of magnitude \cite{Persson89}. This result led to a conclusion that
other mechanisms, such as electron-hole pair excitation, must be 
dominant in the decay of Pt--CO vibrations.

Our suggested value of the off-diagonal force constant $k_1$ is 
large enough to completely change this situation. When this term
is included, we obtain many phonon modes $v_i$ with couplings $h_i$.
The frequencies are still discrete and the model is not 
realistic enough to calculate the total decay numerically. We
therefore just calculate the enhancement of the ratio
\begin{equation}
\sum_{i} \left({h_i \over g}\right)^2
\end{equation}
with respect to the case with $k_1=0$, which turns out to be 44.4.
The total enhancement of the decay probability via three-phonon 
processes is the third power of this value, or $0.88\times10^5$, which 
is again about the right magnitude as the experimental values,
according to the estimates of Ref. \cite{Persson89}.

Thus we find that the three-phonon decay of the Pt--CO vibrations
for a single adsorbate agrees with the observed linewidth, and no
other decay processes (or other mechanisms which increase the width)
need to be involved. Furthermore, the phonon mechanism can explain
the observed coverage dependence of the linewidth, as discussed in
Ref. \cite{Brako95}. The efficiency of the decay depends upon the
ability of the surrounding Pt atoms to move laterally following
the Pt--CO vibration, in much the same way as their ability to have 
a {\it static} relaxation shift determines the contribution to the
adsorption energy. In infrared spectroscopy experiments, CO atoms
move in phase over very large distances, and in dense adsorbate
structures there are few or no atoms which can move laterally and
contribute to the decay of the Pt--CO vibrations. This is in 
agreement with the observed minimum of the adsorption width at a
coverage of 0.3 \cite{Persson89}, whereas the width is large at a
coverage of 0.19, corresponding to a less dense regular structure.
If the width were due to inhomogeneous effects, one would expect
the converse.

To conclude, we have shown that the interaction between 
CO molecules adsorbed in on-top positions 
on the Pt(111) surface is associated with the
lateral relaxation of the surrounding Pt atoms,
and that a  good quantitative
description is obtained by choosing an appropriate value of the relaxation. 
The model is further corroborated by the fact that it gives 
a value for the multiphonon decay of Pt--CO vibrations which is 
in agreement with the vibrational linewidth observed experimentally. 
This approach predicts that the interaction between equivalent species
will usually be repulsive, since some surface atoms will not be able to relax
completely. This is indeed the case for on-top CO on Pt(111) considered
here \cite{Skelton94}, and also for CO
in three-fold hollow sites on Ni(111) \cite{Skelton97}.
Non-equivalent species, however, may interact attractively if
the relaxation forces act in the same direction on some surrounding
surface atoms, increasing the total relaxation energy.
This appears to be the case between coadsorbed CO in on-top and NO in
bridge positions on Pt(111), and presumably also between on-top and
bridge CO which occur on Pt(111) at higher coverages. We
intend to pursue a quantitative investigation of these systems.

\ack{This work was supported by the Ministry of Science and
Technology of the Republic of Croatia under the contract
Nr. 00980102.}

\end{document}